# A new semiconducting perovskite alloy system made possible by gas-source molecular beam epitaxy

Ida Sadeghi[1], Jack Van Sambeek[1], Tigran Simonian[2], Michael Xu[1], Kevin Ye[1], Valeria Nicolosi[2], James M. LeBeau[1], R. Jaramillo[1†]

1. Department of Materials Science and Engineering, Massachusetts Institute of Technology, Cambridge, MA, USA

2. School of Chemistry & CRANN, Trinity College Dublin, College Green, Dublin, Ireland

† Corresponding author: rjaramil@mit.edu

**Optoelectronic technologies are based on families of semiconductor alloys. It is rare that a new semiconductor alloy family is developed to the point where epitaxial growth is possible; since the 1950s, this has happened approximately once per decade. Here we demonstrate epitaxial thin film growth of semiconducting chalcogenide perovskite alloys in the Ba-Zr-S-Se system by gas-source molecular beam epitaxy (MBE). We stabilize the full range $y = 0 \ldots 3$ of compositions $BaZrS_{(3-y)}Se_y$ in the perovskite structure, up to and including $BaZrSe_3$, by growing on $BaZrS_3$ epitaxial templates. The resulting films are environmentally stable and the direct band gap ($E_g$) varies strongly with Se content, as predicted by theory, covering the range $E_g = 1.9 \ldots 1.4$ eV for $y = 0 \ldots 3$. This creates possibilities for visible and near-infrared (VIS-NIR) optoelectronics, solid state lighting, and solar cells using chalcogenide perovskites.**

Each new family of semiconductors expands opportunities for technological innovation, as new property combinations become available. Alloy families are selected based on their different properties for different applications, from cameras, to solar cells, to telecommunications, and so on. Device engineering and optimization are then achieved through the fine control of composition and electronic structure made possible by epitaxial film growth. Chalcogenide perovskites are notable for their high dielectric polarizability, chemical and environmental stability, and wide range of theoretically-predicted $E_g$.[1]

By perovskite, we refer to phases with composition $ABX_3$, with structure characterized by corner-sharing $BX_6$ octahedra (including distorted perovskites, with $\angle(B-X-B) \neq 180°$). Most $ABX_3$ chalcogenides with $X =$ S, Se are thermodynamically unstable in the perovskite structure, and instead form edge- and face-sharing structures with needle-like crystals.[2] The most-studied chalcogenide perovskite is $BaZrS_3$ (distorted perovskite structure, orthorhombic space group $Pnma$). It can be prepared by high-temperature bulk and thin film growth methods, including MBE, and it is very stable, even against heating in air above 500 °C.[3,4,5] It has a direct band gap of $E_g = 1.9$ eV and absorbs light strongly.[3] For VIS-NIR optoelectronic applications, solid state lighting, and solar energy conversion, a range of $E_g$ is necessary. Theory predicts that Se-alloying rapidly decreases $E_g$, to as low as $\approx 1.3$ eV in the pure selenide.[6,7,8] This would make $BaZrS_{(3-y)}Se_y$ perovskite alloys particularly attractive for thin-film solar cells. However, $BaZrSe_3$ is not stable in the perovskite structure. Theory predicts that the edge-sharing, needle-like structure with $E_g \approx 1$ eV is the most stable, whereas synthesis attempts yielded a hexagonal, defect-ordered phase with metallic conductivity.[6,8–10] Powder synthesis results demonstrate that BaZrS$_{(3-y)}$Se$_y$ is stable in the bulk up to at least $y = 1.2$ (40% Se).[11]

The gas-source MBE method, that we introduced recently for making BaZrS$_3$ epitaxial thin films, is suited for expanding into alloy development. We use varying quantities of H$_2$S and H$_2$Se gas flows to make films with varying S:Se ratio, up to and including the selenide perovskite BaZrSe$_3$ with $E_g = 1.4 \pm 0.1$ eV. Our approach represents a revival of methods for making epitaxial films of chalcogenide compound semiconductors by gas-source MBE, developed decades ago for II-VI semiconductors. These methods are newly relevant because of heightened interest in semiconductor fabrication: for chalcogenide perovskites, two-dimensional materials, and other materials for computing, optoelectronics, and energy conversion.

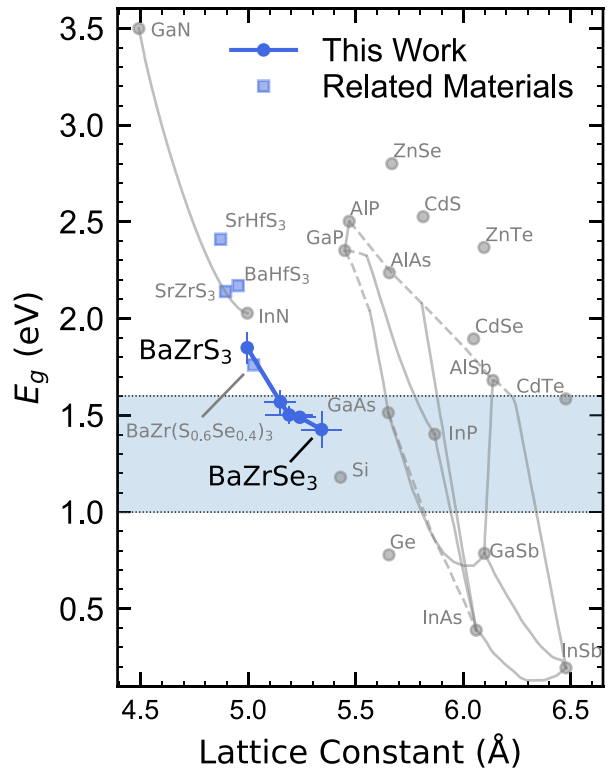

**Figure 1:** Cho plot representing chalcogenide perovskites as a new family of semiconductor alloys. Deep blue points represent the BaZr(S,Se)$_3$ alloy series reported here as epitaxial thin films. Light blue points represent other chalcogenide perovskites, reported elsewhere as solid-state synthesis of bulk powders. Other data are established semiconductors and alloy families. Direct (indirect) band gap alloys are indicated by solid (dashed) lines. The colored band indicates the range of $E_g$ appropriate for single-junction solar cells. For the chalcogenide perovskites, the In **Fig. 1** we show the band gap ($E_g$) and lattice constant measured on five BaZrS$_{(3-y)}$Se$_y$ films spanning $y$ = 0 … 3 and $a_{PC}$ = 4.996 ± 0.004 … 5.34 ± 0.09 Å, along with data for other semiconducting materials. For the chalcogenide perovskites, we plot the pseudo-cubic lattice constant ($a_{PC}$), determined from X-ray diffraction (XRD) measurements of the (202) reflection, and $E_g$ is determined from photoconductivity spectroscopy (PCS). Plots such as these of $E_g$ vs. lattice constant (called a "Cho plot", after the developer of MBE) are the starting point for designing optoelectronics, solid-state lighting, and solar cells. The advent of epitaxial growth of BaZrS$_{(3-y)}$Se$_y$ alloys with tunable $E_g$ means that chalcogenide perovskites may reasonably be placed on a Cho plot, alongside well-established families of semiconductors. Se alloying is effective at reducing $E_g$ into the range needed for single-junction solar cells. $E_g$ may also be raised above 2 eV, into the green lighting region, in compounds containing Sr and/or Hf.[11,12] To-date such materials have been reported only as powders made by solid-state synthesis; alloys and thin-film epitaxy have not been demonstrated. Nevertheless, we indicate these materials in **Fig. 1**, to suggest that the chalcogenide perovskite alloy family may in the future cover a wide range of $E_g$.[1]

We grow films by pseudomorphic epitaxy on BaZrS$_3$ template layers, leveraging our previously-reported process for epitaxial growth of BaZrS$_3$ on LaAlO$_3$ (LAO) substrates.[3] Due to the high growth temperature, there is substantial intermixing of S and Se between the growing BaZrS$_{(3-y)}$Se$_y$ film and the underlying BaZrS$_3$ layer. As a result, the final composition throughout the entire film (including the original template layer) is closely determined by the ratio of H$_2$Se and H$_2$S gases in the vapor phase during the alloy growth stage. This facile anion exchange process is why we refer to the initial BaZrS$_3$ film as a template layer, instead of a buffer layer. We use the relative concentrations of H$_2$Se and H$_2$S in the vapor phase to determine the nominal composition parameter *y*. As shown below, this nominal composition parameter closely predicts the actual film composition, as measured by scanning transmission electron microscopy energy dispersive X-ray spectroscopy (STEM-EDX). Reflection high-energy electron diffraction (RHEED) data acquired during growth shows evidence of single crystalline, perovskite epitaxial films. In **Fig. 2a** we present RHEED data measured along the [100]$_{LAO-PC}$ azimuth during growth of BaZrS$_3$, BaZrS$_2$Se ($y$ = 1) and BaZrSe$_3$ ($y$ = 3) (LAO-PC indicates pseudo-cubic indexing of the LAO crystal structure). The RHEED pattern does not change when the growth is switched from BaZrS$_3$ to BaZrS$_{(3-y)}$Se$_y$, indicating epitaxial growth of the same crystal structure albeit different composition (*i.e.*, pseudomorphic heteroepitaxy). The pattern remains streaky, without sign of discrete points, indicating atomically-smooth film growth. X-ray reflectivity (XRR) and atomic force microscopy (AFM) analysis also show that the top surface and the buried interfaces are smooth (Fig. **S1**).

We present in **Fig. 2b** reciprocal space maps (RSMs) of the out-of-plane (202) film reflections. The lattice constant of BaZrS$_{(3-y)}$Se$_y$ expands with $y$, so that the alloy and selenide (202) reflections appear at lower $q_z$ than for the sulfide BaZrS$_3$. The RSMs show the presence of two (202) reflections, indicative of regions in the film with different d-spacings; out-of-plane, high-resolution x-ray diffraction (HRXRD) measurements also showed two (202) peaks (Fig. **S2**). STEM results (discussed below) reveal that this peak splitting derives from film microstructure. In **Fig. 2c** we present RSMs of the (121) reflections, containing both in- and out-of-plane contributions.

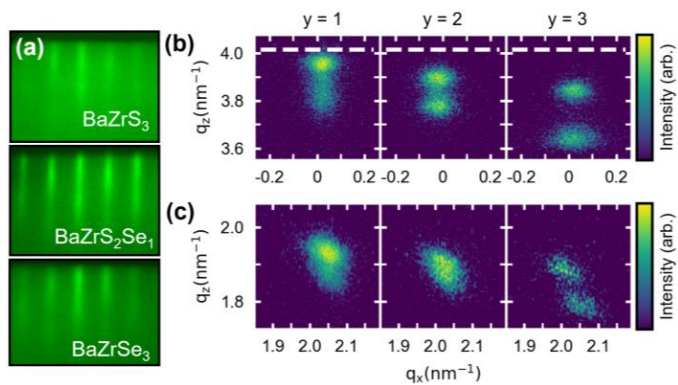

**Figure 2:** Diffraction demonstrates pseudomorphic heteroepitaxy. (a) RHEED data measured along [100] pseudocubic axis of the LAO substate for a BaZrS$_3$ template layer on LAO, a BaZrS$_2$Se film, and a BaZrSe$_3$ film. The long diffraction streaks indicate atomically-smooth surfaces, and the streak alignments indicate pseudomorphic epitaxial alloy growth on the BaZrS$_3$ template. (b-c) RSMs for BaZrS$_2$Se, BaZrSSe$_2$, and BaZrSe$_3$ films, all on BaZrS$_3$ templates. (b) Out-of-plane (202) reflection; white dashed line shows the position of the relaxed, pure BaZrS$_3$. (c) In-plane (121) reflection.

We use STEM to investigate the local microstructure, crystal structure, and composition for two representative samples BaZrS$_2$Se ($y = 1$) and BaZrSSe$_2$ ($y = 2$). In **Fig. 3a and 3b** we present high angle annular dark-field (HAADF) images, measured along the [010]$_{LAO-PC}$ zone axis, that includes the substrate, template layer, and top layer. The substrate/template interface is abrupt and we observe the presence of two epitaxial growth modes, as reported previously.[3] The dominant epitaxial growth mode for BaZrS$_3$ films on LAO is via a self-assembled, incoherent buffer layer, enabling fully-relaxed films despite the large lattice constant mismatch.[3] The interface between the bottom, template layer (grown as BaZrS$_3$) and the top, alloy layer is coherent. Therefore, during alloy growth the top layer induces tensile strain in the template layer, and the template layer places the top layer under compressive strain.

These strains appear to control the formation of Ruddlesden Popper (RP) faults or antiphase boundaries (APBs). APBs in perovskites manifest as rock-salt-like layers that disrupt the corner-sharing octahedral connectivity, and are related to layered phases such as the Ruddlesden-Popper series. APBs can accommodate cation off-stoichiometry and relieve strain.[13,14] We observe APBs in our chalcogenide perovskite epitaxial films, which we ascribe to the difficulty of maintaining exact Ba:Zr = 1:1 stoichiometry during MBE synthesis, without an accessible parameter regime of growth by self-limited adsorption.[15] Here we observe an additional effect: in the bottom layer we observe mostly vertical APBs, whereas in the top layer we observe a high concentration of horizontal APBs. As the Ba:Zr supply rates are unchanged throughout, we hypothesize that this change in the APB profile is related to strain relaxation. The vertical APBs in the bottom help to relieve the tensile strain imposed by the overlying film. By the same logic, as the top overlayer is under compressive strain from the underlying bottom, there is a preference for APBs to orient horizontally. The APB profile illustrates clearly the location of the bottom/top interface, despite chalcogen intermixing. This is sensible, as the diffusivity of two-dimensional extended defects is infinitesimal compared to that of individual chalcogen ions: the APBs nucleate and grow simultaneously with ad-atom attachment, while $S^{2-}$ and $Se^{2-}$ ions may migrate throughout the whole time of film growth.

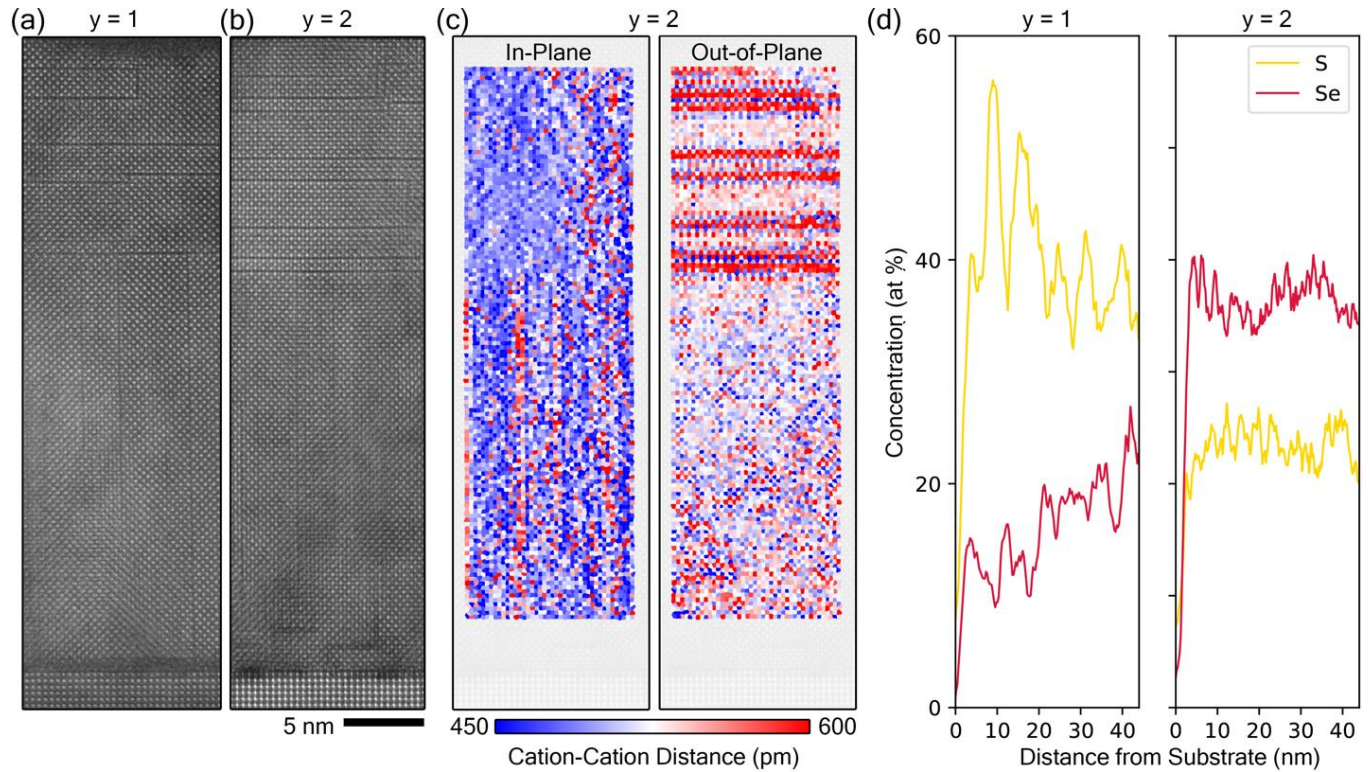

**Figure 3:** STEM measurements of structure and composition. HAADF-STEM image of a typical film cross section of (a) BaZrS$_2$Se ($y = 1$) sample and (b) BaZrSSe$_2$ ($y = 2$). The interface between the bottom layer and the top layer is visible as a change in the distribution of vertical and horizontal APBs. (c) In-plane (left) and out-of-plane (right) cation-cation distances corresponding to the HAADF-STEM image in (b). (d) S and Se atomic concentrations, determined by EDX, through the film thickness for both films.

The APBs locally modulate the d-spacing, and explain the reflection pairs observed in XRD. In **Fig. 3c** we present in-plane and out-of-plane cation-cation distances for sample BaZrSSe$_2$ ($y = 2$) determined by STEM.[16] These are the average of Ba-Ba and Zr-Zr distances; data for sample BaZrS$_2$Se ($y = 1$) and additional analysis are presented in **Fig. S3**. Horizontal APBs create a tripartite population of out-of-plane cation-cation distances in the top layer, corresponding to expansion in the APB core, compression immediately adjacent to the APB core, and an intermediate value away from APBs. The net effect is that the film has regions of distinct d-spacing, despite nearly-uniform composition throughout, due to the presence of APBs. Histograms of the cation-cation spacing measured by STEM (**Fig. S3**) are in quantitative agreement with the reflection pairs observed in RSM (**Fig. 2b-c**) and HRXRD (**Fig. S2**).

In **Fig. 3d** we show STEM-EDX depth-dependent composition profiles. Interestingly, the films have a near-uniform composition from the substrate to the top surface, with Se/S ratio close to the gas conditions during alloy growth. The boundary between the bottom, template layer and the top layer is clear in STEM images due to the APB pattern (**Fig. 3a-b**), but is nearly-imperceptible in the composition profiles. We conclude that S and Se readily diffuse and mix at the film growth temperature, resulting in the conversion of the $BaZrS_3$ template layer into a $BaZrS_{(3-y)}Se_y$ alloy, without disrupting the crystal structure or film microstructure.

We use PCS to measure how $E_g$ varies with composition. Optical measurement techniques including spectrophotometry and ellipsometry can accurately measure $E_g$ for uniform thin films, especially for materials with direct band gap and strong light absorption, such as chalcogenide perovskites.[3,11] However, for films with known variation in composition and $E_g$ from top to bottom, optical methods are not advisable. PCS is sensitive to the layer with the smallest $E_g$ in a multilayer stack (this is why external quantum efficiency measurements are often used to measure $E_g$ of the absorber in thin-film solar cells). We fabricated photodetectors by depositing interdigitated Ti/Au electrodes on the film surface, and we measured the spectral responsivity (the ratio of photocurrent to incident light power) using a tunable light source. In **Fig. 4a** we show the normalized responsivity spectra. The responsivity onset tracks the reduction of $E_g$ with increasing $y$. We derive $E_g$ by extrapolating linear fits to the responsivity immediately above the onset energy, and we present the results in **Fig. 4b** alongside published experimental and theoretical values.[3,7,11,17] For the pure sulfide $BaZrS_3$ we find $E_g = 1.82 \pm 0.08$ eV, agreeing with the value of 1.9 eV determined optically.[3] $E_g$ decreases rapidly with Se content, reaching $1.4 \pm 0.1$ eV for $BaZrSe_3$, confirming theoretical predictions of the effect of Se alloying of $BaZrS_3$.[7,17] The absolute photodetector responsivity decreased monotonically by several orders of magnitude with increasing Se content, going from 400 mA/W for $y = 1$ detector, to 0.04 mA/W for the $y = 3$ detector (**Fig. S5**). This reduction in responsivity indicates a decreasing mobility-lifetime product with increasing Se alloying, which will be addressed through defect engineering in future work.

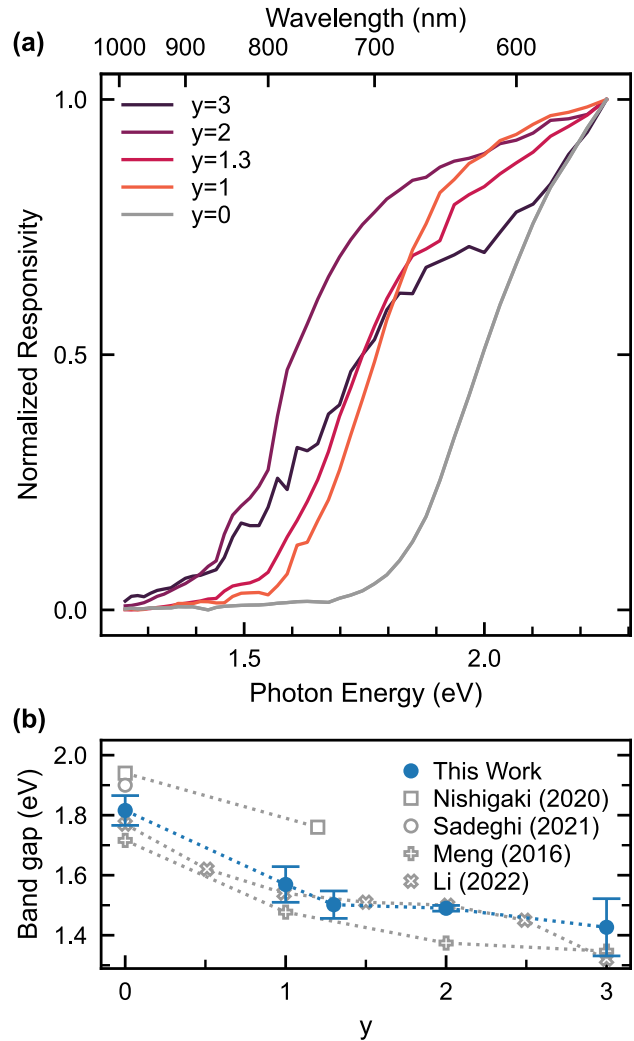

**Figure 4:** Variation of band gap ($E_g$) with Se content, measured by PCS. **(a)** PCS responsivity spectra measured on BaZrS$_{(3-y)}$Se$_y$ photodetectors. **(b)** $E_g$ vs. composition, measured here and as reported elsewhere. Nishigaki (2020): optical measurements on powder samples of BaZrS$_3$ and BaZrS$_{(3-y)}$Se$_y$, $y = 1.2$.[11] Sadeghi (2021): optical measurements on BaZrS$_3$ epitaxial thin films.[3] Meng (2016) and Li (2022) reported theoretical predictions.[7,17]

Our results are notable for demonstrating an alloy family of perovskite semiconductors that can be grown as epitaxial thin films with continuously-variable direct band gap. Chalcogenide perovskites are stable and are made of inexpensive, and low-toxicity elements.[5] Hanzawa *et al.* demonstrated that chalcogenide perovskites can be chemically doped both *n*- and *p*-type, and our results published elsewhere suggest slow excited-state recombination and fast charge mobility.[12,18–20] The phenomenon of anion (S$^{2-}$, Se$^{2-}$) diffusion and mixing observed here suggests that alloy composition and band gap can be controlled by annealing in controlled environments. These reports motivate continued work towards optoelectronic and energy conversion devices based on chalcogenide perovskites, including thin-film solar cells. The most outstanding obstacle to chalcogenide perovskite-based technology may be lowering the temperature required for synthesis of high-quality films.

**Methods**

We grow films using a gas-source MBE system (Mantis Deposition M500). The substrate is heated radiatively from a SiC filament and is rotated at 2 rpm. Ba metal is supplied from an effusion cell (Mantis Comcell 16-500), and Zr metal from an electron beam (e-beam) evaporator (Telemark model 578). We calibrate Ba and Zr source rates using a quartz crystal monitor at the substrate position, and X-ray reflectivity (XRR), X-ray photoelectron spectroscopy (XPS) and X-ray fluorescence (XRF) measurements after film growth. Sulfur and selenium are supplied in the form of $H_2S$ and $H_2Se$ gases (Matheson). These are supplied from condensed, liquified sources of 99.9% and 99.998% purity, respectively, and pass-through point-of-use purifiers (Matheson Purifilter) before entering the MBE chamber. The gases are injected in close proximity to the substrate using custom-made gas lines and nozzles.

We deposit films on $(001)_{PC}$-oriented $LaAlO_3$ (PC stands for pseudo-cubic) single-crystal substrates (CrysTec GmbH). We outgas the substrates in the MBE chamber at 1000 °C in $H_2S$ gas. The growth temperature measured at the thermocouple was 1000 °C. We first grow a template layer of $BaZrS_3$, approximately 20 nm thick. We then interrupt deposition, start the $H_2Se$ flow, adjust the $H_2S$ flow, and resume growth. The $H_2S$ flow rate is 0.8 sccm during substrate outgassing and buffer layer growth. During $BaZrS_{(3-y)}Se_y$ film growth, the $H_2S$ and $H_2Se$ flow rates were adjusted to achieve the desired chalcogen ratio in the gas phase. We measure and control gas flow rates using mass flow controllers (MFCs, Brooks GF100C).

The chalcogen ratio in the gas phase is determined using a residual gas analyzer (RGA, Inficon Transpector 2.0). The $H_2S$ and $H_2Se$ signals appear in the RGA data at 34 Da and 81 Da, respectively (both singly-ionized). We use the relative intensity of these signals to define the composition variable $y$, used throughout this work. The relationship between gas flow rates measured by the mass flow controllers (MFCs) and chalcogen ratio measured by the RGA is nonlinear. To achieve $H_2S/H_2Se = 2$ (composition $y = 1$), the $H_2S$ and $H_2Se$ flow rates were 0.6 sccm and 0.1 sccm, respectively (a ratio of 6). To achieve $H_2S/H_2Se = 0.5$ (composition $y = 2$), the $H_2S$ and $H_2Se$ flow rates were 0.3 and 0.36 sccm, respectively (a ratio of 0.83). For $BaZrSe_3$ ($y = 3$), we use an $H_2Se$ flow of 0.5 sccm and no $H_2S$. The chamber pressure during film growth is approximately $8\times10^{-5}$ torr. The film growth rate is 0.1 Å/s. We maintain $H_2S$ and $H_2Se$ gas flows during cooldown, after growth, to avoid unwanted S and Se desorption. The typical total thickness of the film is approximately 40 nm.

The RGA measurements of $H_2S$ and $H_2Se$ gas-phase concentrations are affected by the relative ionization efficiency of these gases in the RGA. Therefore, the RGA data is likely not an accurate representation of the relative partial pressures in the chamber. $H_2Se$ is less thermodynamically stable than $H_2S$. The observation that the $H_2S/H_2Se$ ratio observed in RGA data is lower than that determined by MFC flow rates may be explained by more effective ionization of $H_2Se$. However, it is interesting to note that the RGA data (that determines *y*) closely tracks the final composition of the films. We suggest that the relative thermodynamic instability of $H_2Se$ affects both ionization efficiency (in the RGA) and solid-vapor reaction rates (at the growing film) to similar extents.

As an alternative to using RGA data, the relative concentrations of $H_2S$ and $H_2Se$ in the chamber can be determined by measuring total chamber pressure with different flow rates, using a cold-cathode vacuum gauge (Inficon Gemini). We find that this approach produces similar results as using the RGA.

We measure RHEED data using a 20 keV, differentially-pumped electron gun (Staib) and a digital acquisition system (k-Space Associates, kSA 400). We perform X-ray reflectivity (XRR) measurements using a Rigaku Smartlab, with a Cu target, and a tube power of 9 kW (45 kV, 200 mA). We perform out-of-plane XRD using a Bruker D8 High-Resolution X-ray diffractometer with a Ge (022) four-bounce monochromator in parallel-beam mode, with a Cu target, and a tube power of 1.6 kW (40 kV, 40 mA). We measured RSMs using a Bruker D8 Discover GADDS with a Co $K_\alpha$ source, ¼ Eulerian cradle, and Vantec-2000 area detector. RSMs were collected with tilts of 0° and 45° for symmetric and asymmetric scans, respectively. The GADDS data were transformed to reciprocal coordinates using Bruker Leptos 7.3 software. Atomic force microscopy (AFM) measurements were performed using a Bruker Icon, and the images were processed using Gwyddion software. PCS was performed on photodetector samples fabricated by sputtering 5 nm Ti/200 nm Au interdigitated contacts on the films. The interdigitated contacts had eight individual fingers, each 5 mm long, with finger spacing and width of 100 µm. We used a tunable light source with a 300 W Xe arc lamp (ScienceTech), providing irradiance between 50 to 200 µW/cm$^2$ depending on the selected wavelength. We used a source-meter (Keithley 2400-C) to source a bias of 4 V and measure the resulting photocurrent.

We prepared samples for STEM using nonaqueous wedge polishing, followed by single-sector ion milling (Fischione 1051) using ion-beam energies of 3, 2, and then 1 kV. We performed STEM imaging on a Thermo-Fischer Scientific Themis Z (probe-corrected, accelerating voltage = 200 kV, convergence angle = 17.9 mrad, beam current = 30 pA). STEM-EDX was performed using the Super X detectors (beam current = 100 pA) and quantified via Thermo-Fisher Scientific Velox software, using both pre- and post-filtering (Gaussian blur, sigma = 1.6). The net intensity profile was processed through a 15 pt adjacent averaging smoothing using Origin software to highlight the trends across the sample. STEM-EELS was performed using a Gatan Continuum spectrometer (sample thickness ≈ 40nm) dispersion = 0.75 eV ch$^{-1}$) and processed via the HyperSpy python package.

**Acknowledgements**

We acknowledge support from the National Science Foundation (NSF) under grant no. 1751736, "CAREER: Fundamentals of Complex Chalcogenide Electronic Materials". A portion of this project was funded by the Skolkovo Institute of Science and Technology as part of the MIT-Skoltech Next Generation Program. K.Y. and J.v.S. acknowledge support by the NSF Graduate Research Fellowship, grant no. 1745302. J.M.L. acknowledges support from the Air Force Office of Scientific Research (FA9550-20-0066). T.S. and V.N. acknowledge support from the SFI Centre for Doctoral Training in Advanced Characterisation of Materials (CDT-ACM) (SFI Award reference 18/EPSRC-CDT/3581). This work was carried out in part through the use of the MIT Materials Research Laboratory and MIT.nano facilities.## References

1. Jaramillo, R. & Ravichandran, J. In praise and in search of highly-polarizable semiconductors: Technological promise and discovery strategies. *APL Mater.* **7**, 100902 (2019).
2. Brehm, J. A., Bennett, J. W., Schoenberg, M. R., Grinberg, I. & Rappe, A. M. The structural diversity of $ABS_3$ compounds with d0 electronic configuration for the B-cation. *J. Chem. Phys.* **140**, 224703 (2014).
3. Sadeghi, I. *et al.* Making $BaZrS_3$ Chalcogenide Perovskite Thin Films by Molecular Beam Epitaxy. *Adv. Funct. Mater.* **31**, 2105563 (2021).
4. Surendran, M. *et al.* Epitaxial Thin Films of a Chalcogenide Perovskite. *Chem. Mater.* **33**, 7457–7464 (2021).
5. Niu, S. *et al.* Thermal stability study of transition metal perovskite sulfides. *J. Mater. Res.* **33**, 4135–4143 (2018).
6. Sun, Y.-Y., Agiorgousis, M. L., Zhang, P. & Zhang, S. Chalcogenide Perovskites for Photovoltaics. *Nano Lett.* **15**, 581–585 (2015).
7. Meng, W. *et al.* Alloying and Defect Control within Chalcogenide Perovskites for Optimized Photovoltaic Application. *Chem. Mater.* **28**, 821–829 (2016).
8. Ong, M., Guzman, D. M., Campbell, Q., Dabo, I. & Jishi, R. A. $BaZrSe_3$: Ab initio study of anion substitution for bandgap tuning in a chalcogenide material. *J. Appl. Phys.* **125**, 235702 (2019).
9. Tranchitella, L. J., Fettinger, J. C., Dorhout, P. K., Van Calcar, P. M. & Eichhorn, B. W. Commensurate Columnar Composite Compounds:  Synthesis and Structure of $Ba_{15}Zr_{14}Se_{42}$ and $Sr_{21}Ti_{19}Se_{57}$. *J. Am. Chem. Soc.* **120**, 7639–7640 (1998).
10. Aslanov, L. A. Selenides of the type $ABSe_3$. *Russ. J. Inorg. Chem.* **9**, 1090–1091 (1964).
11. Nishigaki, Y. *et al.* Extraordinary Strong Band-Edge Absorption in Distorted Chalcogenide Perovskites. *Sol. RRL* **4**, 1900555 (2020).
12. Hanzawa, K., Iimura, S., Hiramatsu, H. & Hosono, H. Material Design of Green-Light-Emitting Semiconductors: Perovskite-Type Sulfide $SrHfS_3$. *J. Am. Chem. Soc.* **141**, 5343–5349 (2019).
13. Jing, H.-M. *et al.* Formation of Ruddlesden–Popper Faults and Their Effect on the Magnetic Properties in $Pr_{0.5}Sr_{0.5}CoO_3$ Thin Films. *ACS Appl. Mater. Interfaces* **10**, 1428–1433 (2018).
14. Bak, J., Bae, H. B., Oh, C., Son, J. & Chung, S.-Y. Effect of Lattice Strain on the Formation of Ruddlesden–Popper Faults in Heteroepitaxial $LaNiO_3$ for Oxygen Evolution Electrocatalysis. *J. Phys. Chem. Lett.* **11**, 7253–7260 (2020).
15. Filippone, S. A., Sun, Y.-Y. & Jaramillo, R. The effect of an improved density functional on the thermodynamics and adsorption-controlled growth windows of chalcogenide perovskites. *MRS Adv.* **3**, 3249–3254 (2018).
16. Aso, R., Kan, D., Shimakawa, Y. & Kurata, H. Atomic level observation of octahedral distortions at the perovskite oxide heterointerface. *Sci. Rep.* **3**, 2214 (2013).
17. Li, Q. *et al.* Control of Polaronic Behavior and Carrier Lifetimes via Metal and Anion Alloying in Chalcogenide Perovskites. *J. Phys. Chem. Lett.* **13**, 4955–4962 (2022).10

18. Niu, S. *et al.* Optimal Bandgap in a 2D Ruddlesden–Popper Perovskite Chalcogenide for Single-Junction Solar Cells. *Chem. Mater.* **30**, 4882–4886 (2018).
19. Ye, K. *et al.* Low-energy electronic structure of perovskite and Ruddlesden-Popper semiconductors in the Ba-Zr-S system probed by bond-selective polarized x-ray absorption spectroscopy, infrared reflectivity, and Raman scattering. *Phys. Rev. B* **105**, 195203 (2022).
20. Ye, K., Zhao, B., Diroll, B. T., Ravichandran, J. & Jaramillo, R. Time-resolved photoluminescence studies of perovskite chalcogenides. *Faraday Discuss.* **239**, 146–159 (2022).
21. Nishigaki, Y. *et al.* Extraordinary Strong Band-Edge Absorption in Distorted Chalcogenide Perovskites. *Sol. RRL* **4**, 1900555 (2020).

**Supplementary information for "A new semiconducting perovskite alloy system made possible by gas-source molecular beam epitaxy"**

Ida Sadeghi[1], Jack Van Sambeek[1], Tigran Simonian[2], Michael Xu[1], Kevin Ye[1], Valeria Nicolosi[2], James M. LeBeau[1], R. Jaramillo[1†]

1. Department of Materials Science and Engineering, Massachusetts Institute of Technology, Cambridge, MA, USA

2. School of Chemistry & CRANN, Trinity College Dublin, College Green, Dublin, Ireland

† Corresponding author: rjaramil@mit.edu

## SUPPLEMENTARY NOTE 1: Smooth surfaces and interfaces

Here, we provide additional evidence (in addition to RHEED in **Fig. 2**, and STEM in **Fig. 3**) as to the surface and film-substrate smoothness. In **Fig. S2a**, we present the result of the XRR measurement on the films. The presence of well-defined Kiessig interference fringes in the XRR curves indicates that the film surface and the film-substrate interface are smooth in all the samples studied. In **Fig. S2b**, we present AFM data measured on the film top surfaces, showing that they are smooth with a r.m.s. roughness between $0.83 \pm 0.27$ and $2.63 \pm 2.34$ nm.

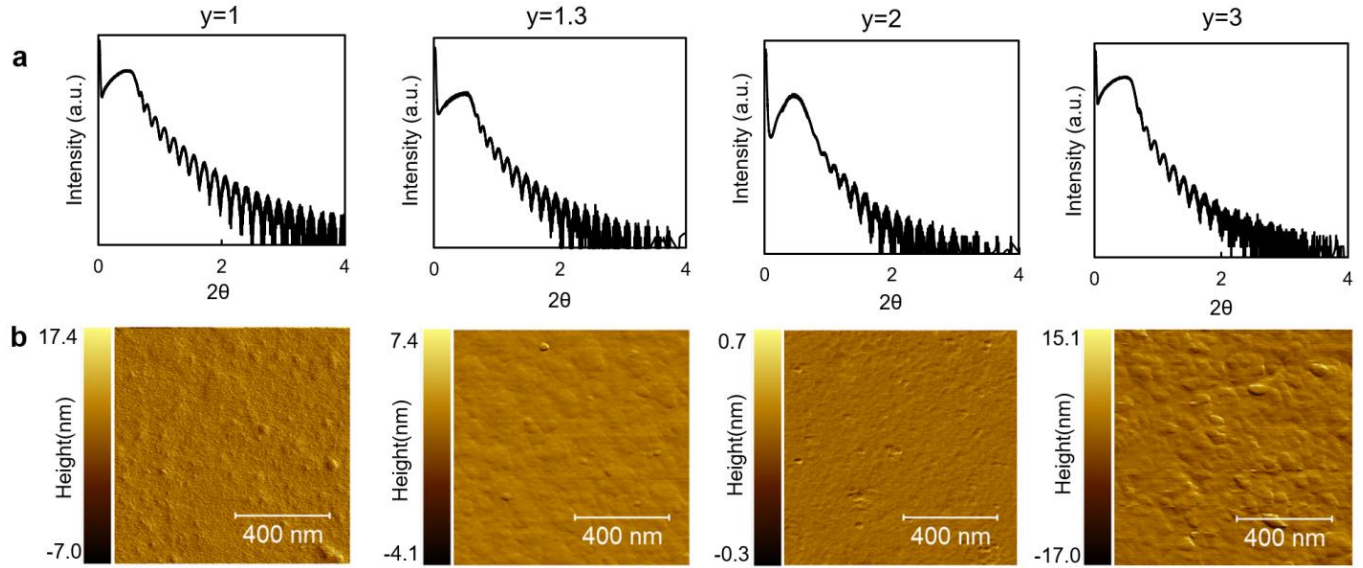

**Figure S1:** (a) XRR and (b) AFM data. The samples have r.m.s. roughness of $2.19 \pm 0.54$ nm ($y = 1$), $1.52 \pm 0.38$ nm ($y = 1.3$), and $0.83 \pm 0.27$ nm ($y = 2$), and $2.63 \pm 2.34$ nm ($y = 3$).

## SUPPLEMENTARY NOTE 2: Out-of-plane X-ray diffraction

We present in **Fig. S2** high resolution XRD (HRXRD) data. The out-of-plane scans for all the films show two (202) reflections corresponding to two sets of out-of-plane d-spacings aligning with the LaAlO$_3$ (012) family of reflections. BaZrSe$_3$ has a larger lattice compared to BaZrS$_3$, therefore, the (202) reflections shift to smaller Bragg angles, *i.e.* larger d-spacing, as the Se to S ratio increases. The pair of reflections results from film microstructure, as described in the main text.

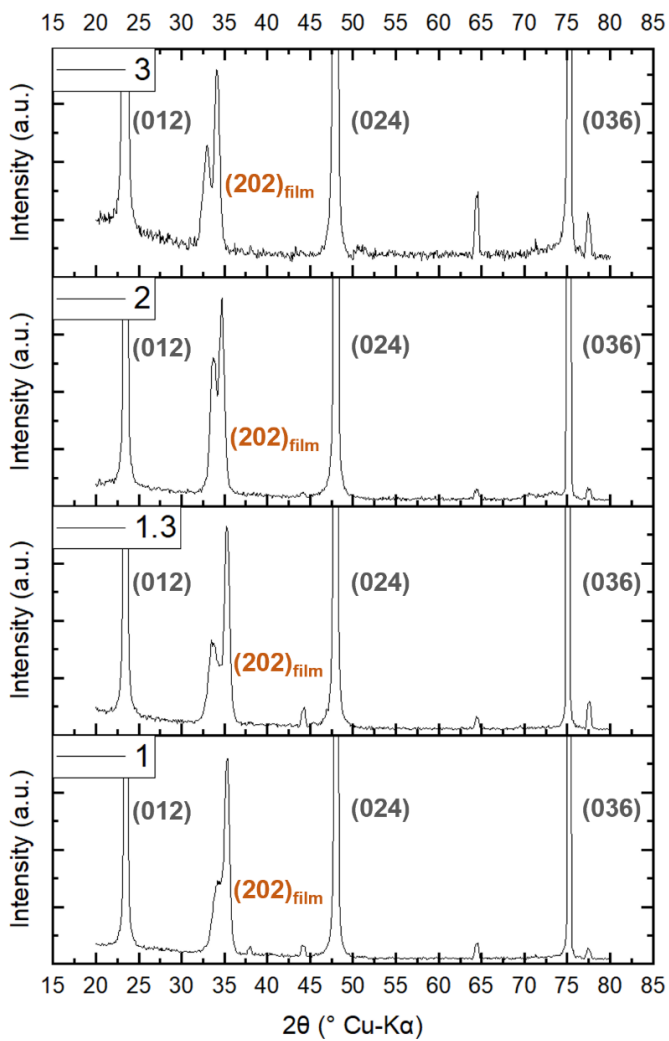

**Figure S2:** HR XRD data of BaZrS$_{(3-y)}$Se$_y$ alloy series ($y$ = 1, 1.3, 2 and 3). The (012), (024) and (036) reflections labeled in grey are from the LaAlO$_3$ substrate.

## SUPPLEMENTARY NOTE 3: Ba-Ba and Zr-Zr spacing maps in the alloy films

We mapped the interatomic cation spacings (averaging both Ba-Ba and Zr-Zr data) in the BaZrS$_2$Se and BaZrSSe$_2$ films. The bottom, template layer that was grown as BaZrS$_3$ shows no significant out-of-plane cation-cation spacing variation. The top layer, where the horizontal anti-phase boundary (APB) layers appear, shows a tripartite distribution of spacings for sample $y$ = 2:

within, immediately adjacent to, and away from the APBs. The regions away from the APBs produce the most intense peak in the histogram, due to their larger total volume. It is this peak (at 533 pm), together with the central peak in the distribution of data from the bottom layer (at 509 pm), that produce the twin peaks in RSM and HRXRD data (**Figs. 2, S2**). The XRD d-spacing data are indicated by dashed lines for comparison. The distribution for sample $y = 1$ is similarly broad, suggestive of multiple peaks. It is not as easy to model as the $y = 2$ case, but it is reasonable to assume that it displays a similar effect of APBs on local structure.

The out-of-plane spacing distributions are broader for the top layers than for the bottom layers, due to the presence of horizontal APBs in the top layers. Similarly, the in-plane spacing distributions are broader for the bottom layers than for the top layers, due to the presence of vertical APBs in the bottom layers.

The presence of APBs and resulting spread in d-spacing complicates a straightforward analysis of the variation in lattice constant measured by XRD with alloy composition. However, in the light of the STEM analysis, we can identify the twin peaks in XRD as representative of regions with high- and low-APB density. In **Fig. S4**, we show that the d-spacing for each of these two populations varies linearly with composition, in accordance with Vegard's law.

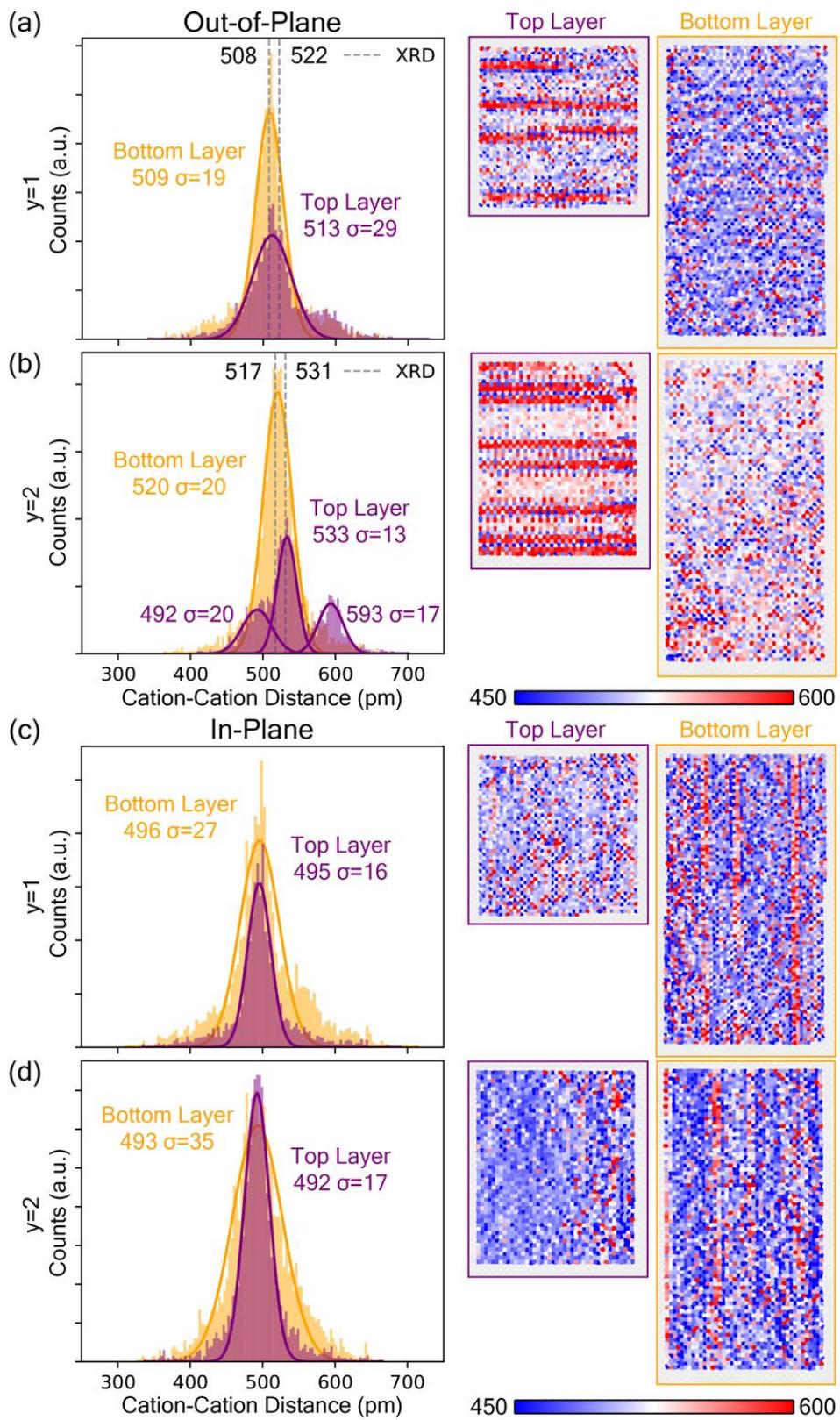

**Figure S3:** Local cation-cation spacing measured by STEM on samples BaZrS$_2$Se ($y = 1$ film) and BaZrSSe$_2$ ($y = 2$). Colormaps show average of Ba-Ba and Zr-Zr spacings (pm). (a) Out-of-plane spacings. In the top layer, the presence of horizontal APBs results in a broad distribution with multiple peaks, corresponding to different regions of the film. In contrast, the spacings in the bottom layer are more narrowly distributed. For the case $y = 2$, a tripartite distribution in the top layer is easily modeled: (i) 593 pm within the APBs, (ii) 492 pm, immediately around the APBs, and (iii) 533 pm, away from the APBs. The dashed lines and accompanying labels indicate the d-spacing of the twin peaks observed in XRD. (b) In-plane spacings. In this case, the distribution is broader for the bottom layer, due to the presence of vertical APBs.

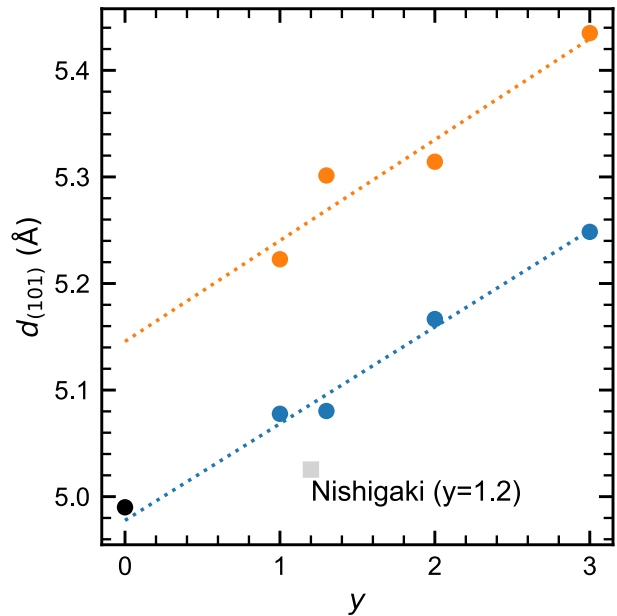

**Figure S4:** $d_{(101)}$ spacing determined from fitting the (202) doublets in **Fig. S2**. The orange points indicate the population of larger $d$-spacings, corresponding to high-APB density material. The blue points indicate the population of smaller $d$-spacings, corresponding to low-APB density material. The dotted lines are linear fits to the $y > 0$ data, in accordance with Vegard's law. The black circle at $y = 0$ represents $d_{(101)}$ for a low-APB-density BaZrS$_3$ film; the grey square indicates data from the literature for a powder pellet of composition $y = 1.2$.[21]

## SUPPLEMENTARY NOTE 4: Composition trend in photodetector responsivity

For the epitaxial photodetectors, the absolute responsivity varies strongly with film composition at all wavelengths. **Fig. S5** shows this responsivity and external quantum efficiency (EQE) trend at 600 nm illumination. The polycrystalline pure sulfide detector shows the lowest responsivity compared to the epitaxial $BaZrS_3$ and all $BaZrS_{(3-y)}Se_y$ detectors. The lower responsivity of the polycrystalline $BaZrS_3$ detector is a result of its very different defect concentrations (including a high density of twin boundaries, shown here). The epitaxial $BaZrS_3$ and alloy devices are made from films with total thickness approximately 40 nm, whereas the polycrystalline $BaZrS_3$ device is made from a film 23 nm thick. Although differences in thickness can affect responsivity and EQE, they do not explain the large differences observed here.

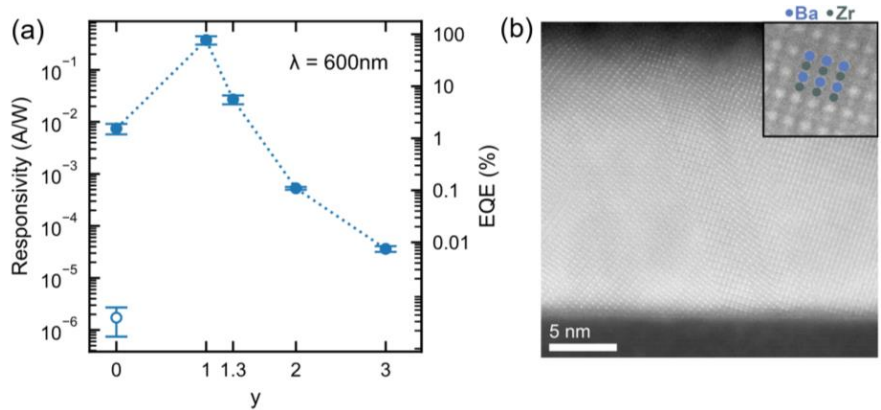

**Figure S5:** (a) Absolute responsivity and EQE of polycrystalline $BaZrS_3$ (hollow circle marker) as well as epitaxial $BaZrS_{(3-y)}Se_y$ (solid circles markers) photodetectors at illumination wavelength of 600 nm for varying composition. (b) STEM HAADF image along the film $[110]_{PC}$ zone of the non-epitaxial polycrystalline pure-sulfide film deposited on sapphire, with inset showing the corresponding Ba and Zr atom columns. Several crystal orientations are visible, with many twin boundaries present.

## SUPPLEMENTARY NOTE 5: Table of samples reported here

The below table records the samples reported here, and their names as appearing in our laboratory notebooks and database.

| Sample description | Database name |
|---|---|
| polycrystalline $BaZrS_3$ | G018 |
| $BaZrS_3$ ($y=0$) | G039 |
| $BaZrS_{(3-y)}Se_y$ ($y=1$) | G069 |
| $BaZrS_{(3-y)}Se_y$ ($y=1.3$) | G066 |
| $BaZrS_{(3-y)}Se_y$ ($y=2$) | G065 |
| $BaZrS_{(3-y)}Se_y$ ($y=3$) | G060 |